\documentclass[aps,pra,twocolumn,floatfix]{revtex4}
\usepackage{bm}
\usepackage{epsfig}

\begin{document}

\title{A simple and surprisingly accurate approach to the chemical bond obtained
from dimensional scaling}
\author{Anatoly A. Svidzinsky$^{a,b}$, Marlan O. Scully$^{a,b,c}$ and Dudley R.
Herschbach$^d$}
\affiliation{
$^a$Depts. of Chemistry, and Mechanical and Aerospace Engineering,
Princeton University, Princeton, NJ 08544\\
$^b$Depts. of Physics, Chemical and Electrical Engineering,
Texas A\&M University, TX 77843-4242 \\
$^c$Max-Planck-Institut f\"ur Quantenoptik, D-85748 Garching, Germany\\
$^d$Department of Chemistry and Chemical Biology,
Harvard University, Cambridge, MA 02138
}
\date{\today }

\begin{abstract}
We present a new dimensional scaling transformation of the Schr\"odinger
equation for the two electron bond. This yields, for the first time, a good
description of the two electron bond via D-scaling. There also emerges, in
the large-D limit, an intuitively appealing semiclassical picture, akin to a
molecular model proposed by Niels Bohr in 1913. In this limit, the electrons
are confined to specific orbits in the scaled space, yet the uncertainty
principle is maintained because the scaling leaves invariant the
position-momentum commutator. A first-order perturbation correction,
proportional to 1/D, substantially improves the agreement with the exact
ground state potential energy curve. The present treatment is very simple
mathematically, yet provides a strikingly accurate description of the
potential energy curves for the lowest singlet, triplet and excited states
of H$_2$. We find the modified D-scaling method also gives good results for
other molecules. It can be combined advantageously with
Hartree-Fock and other conventional methods.
\end{abstract}

\maketitle

Quantum chemistry has achieved excellent agreement between theory and
experiment, even for large molecules, by using computational power to
overcome the difficulty of treating electron-electron interactions \cite
{Scha84,Parr,Scul02,Berr82}. Here we present a new version of an
unconventional method to treat electronic structure \cite
{Hers86,Loes87,Hers92,Fran88}. This emulates an approach developed in
quantum chromodynamics \cite{Witt80}, by generalizing the Schr\"odinger
equation to D dimensions and rescaling coordinates \cite{Hers92}.

\begin{figure}
\bigskip
\centerline{\epsfxsize=0.4\textwidth\epsfysize=0.35\textwidth
\epsfbox{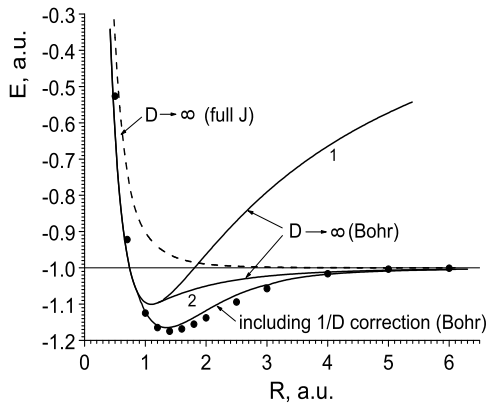}}

\caption{
Energy $E(R)$ of H$_2$ molecule in the limit $D\rightarrow \infty $
calculated from Eq. (\ref{d1})
(dashed curve) and from the Bohr model of Eq. (\ref{a18})
(solid curves). Curve 1 corresponds to a symmetric configuration
obtained by Bohr \cite{Bohr1} and pictured in Fig. \ref{HLD} (top). Curve 2
describes an asymmetric solution (not found by Bohr),
see Fig. \ref{HLD} (bottom).
Lower solid curve is the improved ground state $E(R)$ after including 1/D
correction.
}
\label{h2d}
\end{figure}

Early work found the tutorial D-scaling procedure of Witten \cite{Witt80}
can be dramatically improved; the ground state energy of He was obtained
accurate to 5 significant figures by interpolation between the $D=1$ and $%
D\rightarrow \infty $ limits \cite{Hers86}, and to 9 figures by a
perturbation expansion in 1/D \cite{Good92}. However, the scaling procedure
which worked well for atoms \cite{Hers86,Loes87} did not prove successful
for two-center problems \cite{Hers92,Fran88}; e.g., for H$_2$ that procedure
did not yield a bound ground state, (see our Fig. \ref{h2d}).

\begin{figure}[h!]

\includegraphics[angle=0,width=8cm]{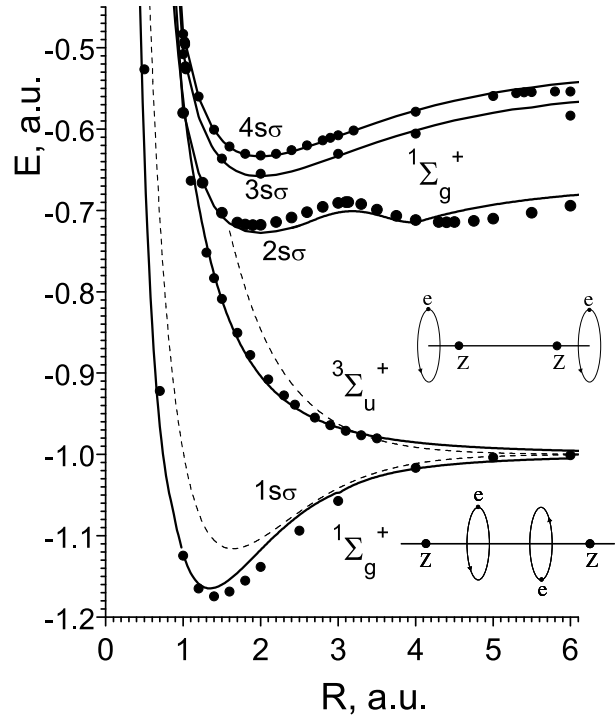}


\caption{
Potential energy (solid curves) of the ground and a few excited states
of H$_2$ obtained from the Bohr model with
D-scaling analysis.
Dots are the ``exact" energies \cite{dot}.
The inserted
figures on the right hand side depict the two nuclei of charge
$Z$
and Bohr's ``planetary" orbits for the electrons in the
$^1\Sigma _g^{+}$ and $^3\Sigma _u^{+}$ states (see also Fig. \ref{HLD}).
Dashed curves are from Heitler-London treatment \cite{Heit27}.
}
\label{h2gexc}
\end{figure}

In our present approach, the large-D limit makes contact with the Bohr model
of the H$_2$ molecule \cite{Bohr1}. In this way we obtain, for the first
time, a link between pre- and post-quantum mechanical descriptions of the
chemical bond (Bohr-Sommerfeld vs Heisenberg-Schr\"odinger). Marked
improvement is achieved by including the leading correction term in $1/D$
and a rudimentary adjustment of the D-scaling. Fig. \ref{h2gexc} shows
potential energy curves for H$_2$ obtained with our simple approach. Dots
comprise a synthesis of experimental data and computations employing many
terms in variational wavefunctions \cite{dot}. Our simple method gives
surprisingly accurate results and holds promise for numerous applications.

\begin{figure}
\bigskip
\centerline{\epsfxsize=0.3\textwidth\epsfysize=0.15\textwidth
\epsfbox{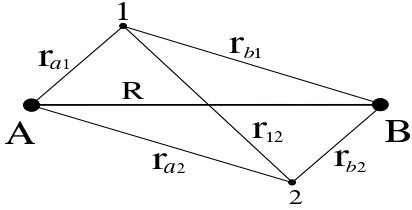}}

\caption{Electronic distances in
H$_2$ molecule. The nuclei A and B are fixed a distance $R$ apart.
}
\label{coord}
\end{figure}

We first outline our method as applied to H$_2$ and then indicate how it
differs from what preceded \cite{Hers92,Fran88}. Fig. \ref{coord} displays
electron distances in the H$_2$ molecule. All distances are expressed in
terms of the Bohr length $a_0=\hbar ^2/me^2$, where $m$ is the electron
mass, and energies are in the Hartree unit $e^2/a_0$. We start with $\hat
H\Psi =E\Psi $, for H$_2$:
$$
\hat H=-\frac 12\nabla _1^2-\frac 12\nabla _2^2+V(\rho _1,\rho
_2,z_1,z_2,\phi ).
$$
The Coulomb potential energy $V$ is given by
\begin{equation}
\label{a16}V=-\frac Z{r_{a1}}-\frac Z{r_{b1}}-\frac Z{r_{a2}}-\frac
Z{r_{b2}}+\frac 1{r_{12}}+\frac{Z^2}R,
\end{equation}
in terms of distances defined in Fig. \ref{coord}. In cylindrical coordinates%
$$
r_{ai}=\sqrt{\rho _i^2+\left( z_i-\frac R2\right) ^2},\quad r_{bi}=\sqrt{%
\rho _i^2+\left( z_i+\frac R2\right) ^2},
$$
$$
r_{12}=\sqrt{(z_1-z_2)^2+\rho _1^2+\rho _2^2-2\rho _1\rho _2\cos \phi },
$$
where $R$ is the internuclear spacing and $\phi $ the dihedral angle between
the planes containing the electrons ($i=1$, $2$) and the internuclear axis.

We proceed by endowing each vector with $D$ cartesian coordinates \cite
{Hers92}. The potential energy $V$ is retained \cite{note} in the three
dimension form of Eq. (\ref{a16}) whereas the Laplacians in the kinetic
energy take the form
\begin{equation}
\label{a3}\nabla ^2=\frac 1{\rho ^{D-2}}\frac \partial {\partial \rho
}\left( \rho ^{D-2}\frac \partial {\partial \rho }\right) +\frac 1{\rho ^2}%
\frac{\partial ^2}{\partial \varphi ^2}+\frac{\partial ^2}{\partial z^2}.
\end{equation}
We then scale coordinates by $f^2$ and energy by $1/f^2$, with $f=(D-1)/2$,
and transform the wavefunction $\Psi $ by
\begin{equation}
\label{a4}\Psi =(\rho _1\rho _2)^{-(D-2)/2}\Phi .
\end{equation}
This recasts the Schr\"odinger equation as
\begin{equation}
\label{a4s}(K_1+K_2+U+V)\Phi =E\Phi ,
\end{equation}
where
$$
K_i=-\frac 2{(D-1)^2}\left\{ \frac{\partial ^2}{\partial \rho _i^2}+\frac{%
\partial ^2}{\partial z_i^2}+\frac 1{\rho _i^2}\frac{\partial ^2}{\partial
\phi ^2}\right\} ,
$$
$i=1,$ $2$ and
\begin{equation}
\label{a4s1}U=\frac{(D-2)(D-4)}{2(D-1)^2}\left( \frac 1{\rho _1^2}+\frac
1{\rho _2^2}\right) .
\end{equation}
In the limit $D\rightarrow \infty $ the derivative terms in $K_i$ are
quenched. The corresponding energy $E_\infty $ for any given internuclear
distance $R$ is then obtained simply as the extremum of the effective
potential, $U+V$, given by
\begin{equation}
\label{a18}E=\frac 12\left( \frac 1{\rho _1^2}+\frac 1{\rho _2^2}\right)
+V(\rho _1,\rho _2,z_1,z_2,\phi ,R).
\end{equation}
This is exactly the energy function that applies to the Bohr model of the
molecule \cite{Bohr1}.

The usual D-scaling procedure \cite{Hers92,Fran88} involves setting up the
full Laplacian in D-dimension and transforming the wavefunction by
incorporating the square root of the Jacobian via $\Psi \rightarrow
J^{-1/2}\Phi $, where $J=(\rho _1\rho _2)^{D-2}(\sin \phi )^{D-3}$. Then, on
scaling the coordinates by $f^2$ and the energy by $1/f^2$, the
Schr\"odinger equation in the limit $D\rightarrow \infty $ yields
\begin{equation}
\label{d1}E=\frac 12\left( \frac 1{\rho _1^2}+\frac 1{\rho _2^2}\right)
\frac 1{\sin ^2\phi }+V(\rho _1,\rho _2,z_1,z_2,\phi ,R),
\end{equation}
which differs from Eq. (\ref{a18}) by the factor $1/\sin ^2\phi $.

Our procedure, designed to reduce to the Bohr model at the large-$D$ limit,
instead incorporates only the radial portion of the Jacobian in transforming
the wavefunction via Eq. (\ref{a4}). This has important consequences. Fig.
\ref{h2d} displays the $D\rightarrow \infty $ potential energy curve of Eq. (%
\ref{d1}) (dashed curve, ``full-$J$''); which exhibits no binding. However,
our ``Bohr model'' limit obtained from Eq. (\ref{a18}) yields a good
zero-order approximation for the ground state (curve 2 in Fig. \ref{h2d}).
It is surprisingly accurate at both large and small internuclear distances $%
R $. Also, the model predicts the ground state is bound with an equilibrium
separation $R_e=8/(9-\sqrt{3})\approx 1.10$ and gives the binding energy as $%
E_B=3(2-\sqrt{3})/8$ a.u.$\approx 0.100$ a.u.$=2.73$ eV. The Heitler-London
calculation (shown in Fig. \ref{h2gexc}, dashed curve), obtained from a
two-term variational function, gives $R_e=1.51$ and $E_B=3.14$ eV \cite
{Heit27}, whereas the ``exact'' results are $R_e=1.401$ and $E_B=4.745$ eV
\cite{Scha84}.

For the triplet $^3\Sigma _u^{+}$ state, as seen in Fig. \ref{h2gexc}, the
Bohr model energy function of Eq. (\ref{a18}) gives a remarkably close
agreement with the ``exact'' potential curve and is in fact much better than
the Heitler-London result (which, e.g., is 30\% high at $R=2$).

In essence, D-scaling procedures resemble gauge transformations. Many
varieties of scaling are feasible, subject only to the constraint that as $%
D\rightarrow 3$ the scaled Schr\"odinger equation reduces to the correct
form. The basic aim is to devise a scaling that removes the major, generic
D-dependence, enabling the easily evaluated $D\rightarrow \infty $ limit to
approximate the $D=3$ energy. With the ``full-J'' scaling previously used
\cite{Fran88}, when $D$ is increased the $(\sin \phi )^{D-3}$ factor in the
Jacobian forces $\phi $ towards $90^{\circ }$, while minimization of
electron-electron repulsion requires $\phi \rightarrow 180^{\circ }$. The
effect is to overweight electron repulsion; this is the chief source of the
failure to obtain chemical bonding in previous work. Our new procedure
avoids such overweighting by retaining the $D=3$ form for the $\phi $-part
of both the Jacobian and the Laplacian of Eq. (\ref{a3}). Thereby $\phi $
remains a fully quantum variable as $D\rightarrow \infty $, rather then
being converted to a semiclassical parameter along with the $\rho $ and $z$
coordinates. This much improves description of the electron repulsion and
hence the chemical bonding.

The scaling procedure enables, in the large-D limit, calculations to be
carried out in the scaled space that are entirely classical. The extremum
equations $\partial E/\partial z=0$ and $\partial E/\partial \rho =0$ are
equivalent to Newton's second law applied to the motion of each electron.
Respectively, they specify that the net Coulomb force on the electron along
the $z-$axis vanishes and that the projection of the Coulomb force
perpendicular to the molecular axis balances the centrifugal force. Although
the electrons are thereby confined to specific orbits in the scaled space,
the uncertainty principle is nonetheless satisfied. This is so because the
conjugate momenta are scaled inversely to the coordinates, leaving the
position-momentum commutator invariant. The continuous transition between
the scaled space and the unscaled space in effect relates classical
trajectories at large-D to corresponding quantum distributions at D=3. This
aspect becomes particularly evident when treating electronic tunneling \cite
{Hers92}.

\begin{figure}[!]

\includegraphics[angle=270,width=7cm]{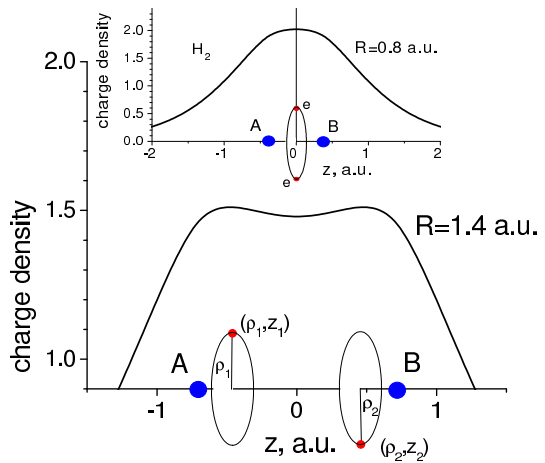}


\caption{Distribution of the electron charge density in the H$_2$ molecule
along the molecular axis $z$. The nuclei are fixed a distance $R$ apart.
Circles are electron orbits in Bohr's model.
}
\label{HLD}
\end{figure}

Fig. \ref{HLD} displays the ``exact'' electron charge density along the
molecular axis in the ground state of H$_2$ for internuclear spacing $R=0.8$
and $1.4$ a.u. Circles show electron orbits in Bohr's model. The orbit
positions for any $R$ actually coincide with the maxima in the charge
density. This provides a link between the wave mechanical and Bohr ($%
D\rightarrow \infty $ limit) treatments of the H$_2$ bond.

The ground state $E(R)$ can be substantially improved by use of a
perturbation expansion in powers of $1/D$, developed by expanding the
effective potential of Eq. (\ref{a18}) in powers of the displacement from
the minimum \cite{Hers92}; for He this has yielded highly accurate results
\cite{Good92}. Terms quadratic in the displacement describe harmonic
oscillations about the minimum and give a $1/D$ correction to the energy. A
symmetry breaking point occurs at $R_c=1.2$, beyond which the electron
orbits move apart (c.f. Fig. \ref{HLD}). Such symmetry breaking is a typical
feature exhibited as $Z$ or $R$ is varied at large-$D$ \cite{Hers92,Shi01}.
The $1/D$ correction works well at points substantially below or above $R_c$%
. Results for those regions thus can be combined. This involves transforming
the axial coordinates to $z_1\pm z_2$, in order to separate the double-well
structure that occurs in $z_1-z_2$. With the other coordinates fixed at
their values at the minimum of $U+V$, a one-dimensional Schr\"odinger
equation is solved to take into account the double-well mode. This
contribution to the $1/D$ correction corresponds to electron resonance or
exchange. The result gives good agreement with the ``exact'' $E(R)$ over the
full range of $R$ (lower solid line in Fig. \ref{h2d}). The $1/D$ correction
predicts the equilibrium separation to be $R_e=1.38$ with binding energy $%
E_B=4.50$ eV.

The Bohr and D-scaling techniques taken together hold promise for numerous
applications. In particular, these provide a new approach to treating
excited states. For example, in our analysis the energy of 1s2s state of the
He atom is obtained as an extremum of the energy function $%
E=n_1^2/2r_1^2+n_2^2/2r_2^2+V({\bf r_1},{\bf r_2})$, where $n_1=1$ and $n_2=2
$; ${\bf r_1}$, ${\bf r_2}$ are electron radius vectors and $V$ is the
Coulomb potential energy. This yields the value of $-2.159$ a.u. which
differs by 0.7\% from the ``exact'' 1s2s energy of $-2.144$ a.u. For other
excited states of He as well as more complex atoms the combination of the
Bohr and D-scaling approaches also provides accurate results; we will
discuss this elsewhere.

\begin{figure}[!]
\bigskip
\centerline{\epsfxsize=0.42\textwidth\epsfysize=0.38\textwidth
\epsfbox{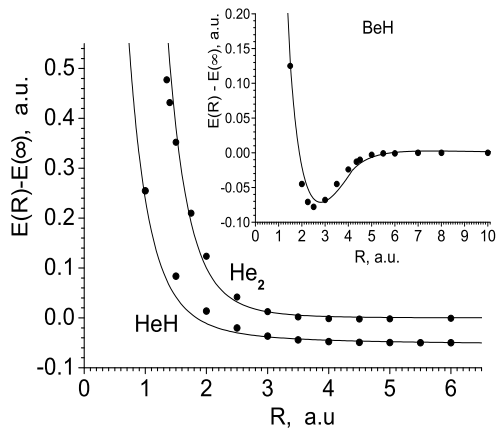}}

\caption{Ground state $E(R)$ of HeH, He$_2$ and BeH molecules
calculated within the Bohr model (solid curves). The HeH curve is shifted
down for clarity.}
\label{HeH}
\end{figure}

Fig. \ref{h2gexc} demonstrates application of our technique to a few excited
states of the H$_2$ molecule. In treating $^1\Sigma _g^{+}$ excited states,
we incorporate D-scaling analysis at large $R$ and the exact $E(R)$ of the H$%
_2^{+}$ molecular ion which provides a good description in the remaining
region. We have also found the present $D\rightarrow \infty $ limit (Bohr
model) gives good results for other molecules; examples so far treated
include HeH, He$_2$, and BeH, pictured in Fig. \ref{HeH}, and LiH, Li$_2$, Be%
$_2$, and the triatomics BeH$_2$ and H$_3$ \cite{H3}.

Another useful strategy is to combine the present approach with conventional
electronic structure methods. At $D=3$, evaluation of the correlation
energy, $E_{\text{corr }}$ (error in the Hartree-Fock approximation) is the
major difficulty. However, at $D\rightarrow \infty $, $E_{\text{corr }}$ can
be evaluated exactly. Results for He and other atoms \cite{Hers92} show that
$E_{\text{corr }}$ for $D\rightarrow \infty $ is smaller than but comparable
to that for $D=3$. For the ground state of H$_2$ we find an accurate energy
curve $E(R)$ can be obtained by adding the $D\rightarrow \infty $
correlation energy to the $E(R)$ given by the Heitler-London effective
charge method. The result is practically identical to the curve obtained
from the $1/D$ correction (Figs. \ref{h2gexc} and \ref{h2d}).

Our modified D-scaling procedure reincarnates the Bohr model. This requires
only elementary concepts and (laptop) computations yet provides a rather
good description of electron-electron interaction and chemical bonding. The
procedure is readily applicable to many-electron molecules, both ground and
excited states. These results encourage efforts to further improve D-scaling
and to augment conventional variational methods for electronic structure to
incorporate the exact correlation energy attainable at the large-D limit.

We wish to thank M. Kim, S. Chin, and G. S\"ussmann for helpful discussions.
This work was supported by the Robert A. Welch Foundation Grant A-1261, ONR,
AFOSR, DARPA and NSF Grant CHE-9986027 (D.R.H).


\begin{thebibliography}{99}
\bibitem{dot}  Throughout this paper ``exact'' energies (solid dots) are
taken from a combination of experimental and theoretical sources. The
theoretical analyses are typically many term (10-100) configuration
interaction computer calculations, see, for example, Ref. \cite{Scha84}.
\bibitem{note}  Changing the Coulombic terms to $r^{-(D-2)}$, corresponding
to solutions of the D-dimensional Laplace equation, would be futile; for
large $D$ the potential energy then becomes extremely short-ranged and bound
electronic states cease to exist.
\end{thebibliography}
\end{document}